\documentclass{article}

\usepackage{arxiv}

\usepackage[utf8]{inputenc} % allow utf-8 input
\usepackage[T1]{fontenc}    % use 8-bit T1 fonts
\usepackage{hyperref}       % hyperlinks
\usepackage{url}            % simple URL typesetting
\usepackage{booktabs}       % professional-quality tables
\usepackage{amsfonts}       % blackboard math symbols
\usepackage{nicefrac}       % compact symbols for 1/2, etc.
\usepackage{microtype}      % microtypography
\usepackage{lipsum}		% Can be removed after putting your text content
\usepackage{graphicx}
\usepackage[round, sort&compress]{natbib}
\usepackage{ulem}

\usepackage{xcolor}
\hypersetup{
    colorlinks,
    linkcolor={red!50!black},
    citecolor={blue!50!black},
    urlcolor={blue!80!black}
}

\newcommand\blfootnote[1]{%
  \begingroup
  \renewcommand\thefootnote{}\footnote{#1}%
  \addtocounter{footnote}{-1}%
  \endgroup
}

\makeatletter
\def\thanks#1{\protected@xdef\@thanks{\@thanks
        \protect\footnotetext{#1}}}
\makeatother

\begin{document}

%\title{Radar-based Nowcasting in Environmental Science and Machine Learning}
\title{A review of radar-based nowcasting of precipitation and applicable machine learning techniques}

\author{Rachel Prudden\textsuperscript{\P \dag \ddag} \thanks{\P \ Corresponding author: rachel.prudden@informaticslab.co.uk} 
	  \And Samantha Adams\textsuperscript{\dag} \thanks{\dag \ Met Office Informatics Lab, Exeter} 
	  \And Dmitry Kangin\textsuperscript{\ddag}  \thanks{\ddag \ University of Exeter}
	  \And Niall Robinson\textsuperscript{\dag \ddag}
	  \And Suman Ravuri\normalfont\textsuperscript{\S} \thanks{\S \ DeepMind, London}
	  \And Shakir Mohamed\normalfont\textsuperscript{\S}
	  \And Alberto Arribas\textsuperscript{\dag \ddag}}

\maketitle

% Uncomment to override  the `A preprint' in the header
\renewcommand{\headeright}{ }

\blfootnote{\\ \tiny{\textit{This work has not yet been peer-reviewed and is provided by the contributing author(s) as a means to ensure timely dissemination of scholarly and technical work on a noncommercial basis. Copyright and all rights therein are maintained by the author(s) or by other copyright owners. It is understood that all persons copying this information will adhere to the terms and constraints invoked by each author's copyright. This work may not be reposted without explicit permission of the copyright owner.\ This work has been submitted to Monthly Weather Review. Copyright in this work may be transferred without further notice.}}}

\begin{abstract}
A `nowcast' is a type of weather forecast which makes predictions in the very short term, typically less than two hours - a period in which traditional numerical weather prediction can be limited. This type of weather prediction has important applications for commercial aviation; public and
outdoor events; and the construction industry, power utilities, and ground
transportation services that conduct much of their work outdoors.  Importantly,
one of the key needs for nowcasting systems is in the provision of accurate
warnings of adverse weather events, such as heavy rain and flooding, for the protection of life and property in such situations. Typical nowcasting approaches are based on simple extrapolation models applied to observations, primarily rainfall radar. In this paper we review existing techniques to radar-based nowcasting from environmental sciences, as well as the statistical approaches that are applicable from the field of machine learning. Nowcasting continues to be an important component of
operational systems and we believe new advances are possible with new partnerships
between the environmental science and machine learning communities.
\end{abstract}

\section{Introduction}

% rewrite
Heavy rainfall events can cause major disruption to human activities. It is desirable to predict these events ahead of time so that decision makers can take action to protect life, property and prosperity. Nowcasting, or short-term forecasting from observations, remains an important tool in predicting these events. 

%We believe there is enormous potential for combining techniques from atmospheric science and machine learning to produce more skilful nowcasts, helping users to make better decisions.

The essential goals of nowcasting are identical to those of all weather forecasting, with the only 
difference being the spatial and temporal scales involved. The World Meteorological Organization \citep{WMOSecretariat2016} distinguishes among the various forecasting time horizons as:

\begin{quote}
``Usually forecasts for the next 0-2 hours are called nowcasting, from 2-12 hours very 
short-range forecasting (VSRF), and short-range forecasting beyond that; but the capabilities of the 
different ranges can vary upon variables and weather situations.'' 
\end{quote}

Radar-based nowcasting emerged in an era of mainly synoptic and mesoscale
weather prediction. Predicting rainfall during that time was a challenge for numerical
weather prediction (NWP) models, since computational
restrictions  limited the resolution at which NWP models could operate. As a result, NWP models were able to capture mesoscale weather patterns such as fronts, but not the smaller-scale
convective patterns that occur within mesoscale systems. Thus, these
models had limited utility in predicting rainfall in the early hours of the
forecast because of its dependence on the unrepresented small scales.

To improve forecasts at early time steps, techniques based on extrapolating the
current weather from radar or satellite observations - described as nowcasting -
began to be developed. These techniques had the advantage of being able to
represent rainfall at the same resolution as the observations, which was
significantly higher than that used by the numerical models. This led to
significant increases in forecast skill in the early hours of the forecast.
The extrapolation models were based on advecting present rainfall observations along their current trajectories, and are described as \textit{advective nowcasting}. Because these models were very simplistic compared to full
NWP, their skill reduced far more quickly.
Nonetheless, \citet{Browning1980ReviewForecasting} estimated that extrapolation algorithms
outperformed NWP models  at forecast times up to around six hours, making them
an essential tool in operational centers.

Since the 1980s, operational NWP performance has seen major technical improvement from two directions. First, model resolution has increased from 10-20 km grid cells to around 1 km, a scale the model can begin to resolve convective processes. Alongside this, improvements in data assimilation (DA) have made it possible to utilise high density radar and satellite observations, leading to more accurate model initialisation. 

However, despite these advances, making optimal use of the information in high-resolution observations has proved a formidable challenge. In many ways convective scale NWP is still in its infancy, as many sources of predictability and simplifying balance relations that are present at synoptic scales no longer hold.
A number of active research areas are summarised in \cite{JuanzhenSunUseChallenges}, and \cite{ScientificPrediction}.
%One specific example of relevance to nowcasting is the NWP background state being inconsistent with newly observed rainfall, leading to spuriously high rainfall predictions at early lead-times as the model attempts to restore balance \citep{BannisterAssimilation}.
The upshot is that while NWP models undoubtedly add skill on synoptic timescales, this is not the case for the small spatial scales and short time frames relevant to nowcasting.
%\citep{WANG20161}. 
Years of progress in NWP modelling have narrowed this gap, but have not eliminated it. 

%Despite this progress, NWP still faces difficulties in forecasting the earliest time steps. Convective-scale data assimilation remains challenging because of its extreme non-linearity and flow dependence, for which traditional synoptic methods are not well suited; non-linear alternatives do exist, but developing practical implementations is difficult for high-dimensional systems such as the atmosphere \citep{Poterjoy2017Convective-ScaleFilter}. As a result, observations such as radar reflectivity can only be assimilated in an approximate sense, for example by using latent heat nudging \citep{JuanzhenSunUseChallenges}. There are also challenges relating to the model spin-up problem, and fast growth of errors at the convective scale \citep{JuanzhenSunUseChallenges, Han2017AData}. This has the effect of restricting model skill during the early hours of the forecast, leaving scope for improvement using observations-driven methods.

The simpler advective nowcasting models have also seen improvements over the last few
decades,
%. While earlier models were based on pure extrapolation, there has recently been a greater focus on convective initiation and evolution\citep{WMOGuidelines}. This shift has been made possible by 
moving beyond purely advection-based models 
%into more general observations-driven nowcasting, 
by incorporating diagnoses for potential areas of precipitation growth or decay. Such models still lag behind NWP in certain key respects: they lack the ability to model the full atmospheric state and associated interactions, and without a traditional DA step the self-consistency of their predictions cannot be guaranteed.
%such diagnoses with the radar observations cannot be guaranteed. 
Even so, this approach has the potential to extend the utility of observations-driven nowcasting further than advection alone.

In moving towards more general data-driven approaches, machine learning (ML) is a natural ally. Recent years have seen the ascendance of deep learning in particular, a family of ML algorithms which can resolve complex non-linear hierarchies of information. The flexibility and conceptual simplicity of ML make it a promising approach to utilising more diverse sources of information, as well as gleaning more complex effects, making observations-driven nowcasting a more powerful tool.

Realising this potential will be a significant research challenge. Several features of the nowcasting problem distinguish it from more conventional ML tasks: the need for dense pixel-wise prediction; the need to handle underlying structure at multiple spatial and temporal scales; the need to handle extremes and out-of-sample events; an underlying need for flow predictions and handling of regime changes; and the need to make and verify probabilistic predictions. 
For some of these challenges, such as dense prediction, the machine learning community has developed techniques that are promising solutions. For others, however, atmospheric science can help \textit{inform} the next generation of ML approaches, which can be broadly applicable to other fields. Table \ref{tab:areas} highlights key research areas, with pointers to the relevant sections. It also outlines areas of ongoing research, where more work is needed.

\begin{table}[t]
\label{tab:areas}
%\centering
\caption{Key challenges posed by the problem of nowcasting and different approaches taken in 
atmospheric science and machine learning, and where fusions of these approaches can lead to new 
solutions.}
\begin{tabular}{l|l|l}
	\hline
	\textbf{Problem Area} & \textbf{Physical Approaches} & \textbf{ML Approaches}  \\
	\hline \hline 
	Dense prediction & S\ref{advection}& S\ref{dense}\\
	Multiple spatial and temporal scales & S\ref{decomp} & S\ref{dense}\\
	Extremes and out-of-sample events & Ongoing research & Ongoing research  \\
	Flow prediction & S\ref{advection} & S\ref{flow} \\
	Regime change & S\ref{conv_dev} & Ongoing research \\
	Probabilistic predictions & S\ref{prob} & S\ref{uncertain}\\
	\hline
\end{tabular}
\end{table}

There is a demand for new tools that allow us to
assimilate diverse spatial observations, and for decision-support tools that
are fast, scalable and actionable; it is in these settings
%, and by building upon strong body of existing knowledge, 
that we see the opportunity for ML.
% and its role in building scalable and accurate statistical models. 
This review aims to bring the fields of atmospheric
nowcasting and machine learning closer together, providing an
overview of our current state of knowledge and future pathways in precipitation
nowcasting. We begin by detailing the existing approaches to nowcasting using radar
extrapolation, survey the approaches from machine learning that can be applied
to the predictive problems in nowcasting, and conclude with a summary of some of
the pathways for future research.

\section{Persistence-based Nowcasting in Atmospheric Science}
%\section{Radar extrapolation}
\label{advection}

Much of the existing work in precipitation nowcasting attempts to incorporate knowledge of the physics of precipitation into simple models. Some of the best known traditional approaches to nowcasting precipitation, according to 
\citet{Germann2002Scale-DependenceMethodology}, are centred around usage of various inductive biases, including:
\begin{enumerate}
	\item climatological values: means, medians, modes, with variances representing uncertainty 
	\item Eulerian persistence, which predicts most recent observation for future ones
	\item Lagrangian persistence, which assumes the  state  of  each  air  parcel is constant,  and  all  change  is  due  to  parcels  moving  with  the  background  flow (advection)
	\item persistence of convective cells and their properties
\end{enumerate}

The Eulerian persistence is described by the following equation 
\citep{Germann2002Scale-DependenceMethodology}:
\begin{equation}
\hat{\Psi} (t_0 + \tau, x) = \Psi (t_0, x),
\end{equation}
where $\Psi(\cdot)$ is the observed precipitation field, $t_0$ is the forecast initial time, $\tau$ is the time 
difference, and $\hat{\Psi}(\cdot)$ is the forecast precipitation field. Meanwhile, the Lagrangian persistence adds into the equation the displacement vector $\lambda$ 
\citep{Germann2002Scale-DependenceMethodology}:
\begin{equation}
\hat{\Psi} (t_0 + \tau, x) = \Psi (t_0, x-\lambda).
\end{equation}

The Lagrangian persistence assumption is generally the more applicable in short-term rainfall prediction, and is the basis for all current radar extrapolation models. In this section, we describe the structure and implementation of these models in their most basic form, before discussing extensions in sections \ref{prob} and \ref{conv_dev}.

Optical flow algorithms can be divided into two stages. 
The first is to estimate an advection field from two or more radar images. The second stage is to predict future observations using the estimated advection field.
%The first is to derive an advection field from two or more radar images. The second stage is using the advection field to extrapolate an observation forward in time. 
Within this framework, field advection algorithms differ in the implementation of one or both stages.

\subsection{Advection field estimation} \label{est_advect}

The main challenge in making use of the Lagrangian persistence assumption is estimating the displacement vector. The goal is, given a sequence of rainfall fields, to find a motion field $(u, v)$ for which the following equation is satisfied:

\begin{equation}
\frac{d\Psi}{dt} = \frac{\partial \Psi}{\partial t} + 
u \frac{\partial \Psi}{\partial x} + v \frac{\partial \Psi}{\partial y}
= 0, \label{eqn:conservation}
\end{equation}

\noindent or equivalently,

\begin{equation}
\label{adveq}
\frac{\partial \Psi}{\partial t} =
u \frac{\partial \Psi}{\partial x} + v \frac{\partial \Psi}{\partial y},
\end{equation}

\noindent where $\Psi$ is the radar reflectivity or a derived rainfall product 
\citep{Horn1981DeterminingFlow, Pierce2012Nowcasting}. This equation is simply a restatement of the definition of Lagrangian persistence in the form of a differential equation.

A complication is that equation \ref{adveq} is under-specified: many possible motion fields might satisfy the conservation equation, not all of which will be physically plausible. It is therefore necessary to impose  extra conditions in order to make a useful estimation.

The simplest possible choice for constraining the advection field is to assume it is given by a single vector. Early work sought a single displacement vector which maximised the cross-correlation coefficient between sequential precipitation maps. This was done by searching a small set of perturbed vectors surrounding an initial guess, this guess being based on the centre of gravity of each image \citep{Austin1974TheForecasting}, or a previous displacement \citep{Austin1978TheProject}.

%by making an initial guess based on the centre of gravity of each image \citep{Austin1974TheForecasting}, or a previous displacement \citep{Austin1978TheProject}, and then computing the cross-correlation at surrounding displacements to find a maximum value.

While the simple assumption of uniform motion can work well for local analysis, it is unrealistic when 
considering larger areas. An initial attempt to allow for non-uniform motion was made by \cite{Rinehart1978Three-dimensionalRadar}, who divide the full area into a number of smaller blocks and maximise the cross-correlation for each block separately. However, as demonstrated in 
\cite{Tuttle1990DeterminationRadar}, this weaker constraint could lead to inconsistent motion estimates 
and discontinuous advection fields. 

The problem of producing spatially consistent estimates of non-uniform motion can be solved by using a variational approach; that is, encoding additional desired properties of the advection field in a cost function to be minimised. This method is similar to one introduced to the optical flow literature in \cite{Horn1981DeterminingFlow}. Here, the cost function to be minimised is:

$$J(\mathbf{u}) = J_\psi + J_2 = \|\Psi (t_0 + \tau, x) - \Psi (t_0, x-\hat{\alpha}) \|^2 + J_2,$$

\noindent where $J_\psi$ is the residual from the conservation equation (\ref{eqn:conservation}), $\Psi$ the precipitation field, $\hat{\alpha}$ the predicted advection, and $J_2$ is a smoothness penalty. The innovation is the smoothness term $J_2$, which enforces a degree of spatial consistency. Various choices of $J_2$ are possible; \cite{Li1995NowcastingOrography.} use the divergence $\frac{\partial u}{\partial x} + \frac{\partial v}{\partial y}$, while \cite{Bowler2004DevelopmentTechniques} instead use the Laplacian $\nabla^2 \mathbf{V}$.

The variational problem can be solved using various optimisation algorithms, such as successive over-relaxation (SOR) \citep{Li1995NowcastingOrography.} and conjugate-gradient methods \citep{Germann2002Scale-DependenceMethodology}. The full variational problem involves an expensive minimisation, and various tricks have been used to improve convergence. In \cite{Li1995NowcastingOrography.} and \cite{Bowler2004DevelopmentTechniques}, variational 
techniques are used as a form of post-processing for a solution found using a cheaper block-based method. Another option is to solve the full variational problem using a hierarchical approach, where solutions are computed at successively higher resolutions with each field being used as an initial guess for the next \citep{Laroche1994AData,Laroche1995RetrievalsAnalysis,Germann2002Scale-DependenceMethodology}.

\subsection{Numerical advection} \label{num_advect}

Once an advection field has been found, it can be used to propagate radar observations forward in time. This is generally done using a semi-Lagrangian scheme. The basic idea is to divide the advection into short time windows and ``follow the streamlines" of the advection field 
\citep{Germann2002Scale-DependenceMethodology}. %DK
However, working with a discretised field is linked to the challenges of numerical modelling.
%\DK

\paragraph{Semi-Lagrangian scheme}

The semi-Lagrangian scheme is a numerical integration method specifically designed for advection 
problems. Unlike generic solvers (e.g. Euler's method), it makes explicit use of the Lagrangian conservation 
constraint of equation (\ref{eqn:conservation}). This states that the Lagrangian derivative is zero; that is, the 
measured quantity \textit{following an air parcel} is constant:

$$\Psi(t^{n}, x_i) = \Psi(t^{n+1}, x_i + u\Delta t)$$

A natural interpretation is that the solution at time $n$ uniquely determines the solution at time $n+1$ on a grid which has moved with the flow. This interpretation forms the basis of a purely Lagrangian method of solution, as described in \cite{Wiin-Nielsen1959OnForecasting}. However, using an increasingly distorted grid is unworkable in practice. A solution to this, suggested in \cite{Sawyer1963AEquation}, is to repeatedly project the solution back onto the original grid using interpolation. This intermediate method is what is known as the semi-Lagrangian scheme.

\paragraph{Backward and forward schemes}

The most commonly used version of the semi-Lagrangian scheme is the backward one, which comprises the following steps \citep{Diamantakis2013TheChallenges}:
\begin{enumerate}
  \item For each point on the target grid, find the corresponding departure point $\mathbf{r}_d$
  \item Interpolate the previous solution to find the value at $\mathbf{r}_d$
\end{enumerate}

\noindent The first step is usually solved by fixed point iteration on the difference vector $\mathbf{\alpha} = 
\mathbf{r} - \mathbf{r}_d$: that is, by iteratively calculating 

$$\mathbf{\alpha} = \Delta t \mathbf{u} \bigg( t_n, \mathbf{x} - \frac{\mathbf{\alpha}}{2} \bigg) $$

\noindent with 2 or 3 iterations normally considered sufficient 
\citep{Germann2002Scale-DependenceMethodology}. For the second step, a range of interpolation 
methods are possible, with bilinear and bicubic interpolation being common choices 
\citep{Bonaventura2004AnFlows}.

The forward version of the scheme is conceptually similar, except that it follows the air parcels forward in time instead of backward. Its basic steps are as follows \citep{Bowler2004DevelopmentTechniques}:
\begin{enumerate}
\itemsep0em 
  \item For each point on the source grid, find the corresponding destination point $\mathbf{r}$
  \item Apply a spreading kernel to distribute the influence of $\mathbf{r}$ onto the target grid
\end{enumerate}

\noindent The reason for replacing interpolation with a spreading kernel is that the destination points will not in general lie on a grid, and so standard methods such as bilinear interpolation do not apply. The problem then becomes selecting an appropriate kernel shape and lengthscale. 
%(As a side note, Gaussian process regression \citep {mackay1997gaussian} could provide an alternative.)

\paragraph{Numerical diffusion}
As discussed in \cite{Germann2002Scale-DependenceMethodology}, both the forward and backward 
semi-Lagrangian schemes lead to the loss of small-scale features through numerical diffusion. In the 
backward scheme this is due to the interpolation step, while in the forward scheme it is due to the use of kernel spreading.

As mentioned above, there are various possibilities for the interpolation step in the backwards 
scheme. The simplest option of bilinear interpolation has the appealing property of making the 
scheme stable even for very long timesteps \citep{Bonaventura2004AnFlows}. Unfortunately, this 
choice is known to lead to excessive numerical diffusion, meaning that high frequency features are 
lost in the advection. Cubic interpolation significantly reduces this diffusion 
\citep{Bonaventura2004AnFlows}, and is widely used. However, 
\cite{Germann2002Scale-DependenceMethodology} found that even cubic interpolation distorted 
the high frequency spectrum too much for their purposes. This led them to consider a modified 
``interpolate once'' approach: instead of interpolating the field at each timestep, each target grid 
point is traced all the way back to its original point of origin, where a single interpolation is carried out. 
They found that this modified method did indeed reduce artificial diffusion.

The forward scheme was chosen in \cite{Bowler2004DevelopmentTechniques} because, unlike the 
backward scheme, it is guaranteed to be conservative. However, they found that this conservation 
property produces a ``banded'' structure in areas of flow divergence. As noted in 
\cite{Germann2002Scale-DependenceMethodology}, these numerical artefacts can be reduced or 
removed by increasing the kernel lengthscale, but at the expense of degrading small-scale features.

\subsection{Cell-based advection}

An alternative to the standard optical flow approaches for radar extrapolation nowcasting is given by cell-based (or object-oriented) methods. Unlike the full motion fields used in the algorithms above, cell-based methods focus on locating and tracking specific convective cells. This approach has an advantage when a specific severe storm is moving differently to surrounding storms, as it can be more easily identified and more accurately tracked. 

Examples of systems following this approach are numerous, and include TITAN 
\citep{DixonM.andWeiner1993TITAN:Methodology, Munoz2018EnhancedCells}, SCIT 
\citep{Johnson1998TheAlgorithm}, CONRAD \citep{Lang2001CellProducts}, TRACE3D 
\citep{Handwerker2002CellAlgorithm}, TRT \citep{HeringAlessandroMSenesi2005NOWCASTINGDATA} and CELLTRACK \citep{Kyznarova2009CELLTRACKCharacteristics}.

Similar to field-based advection, cell-based methods consist of a flow estimation phase followed by an extrapolation phase. However, the way flow fields are estimated is quite different, consisting of an initial stage of cell identification followed by object tracking. 
Cells are typically identified by selecting radar regions above one (e.g. CONRAD), multiple (e.g. SCIT), or iteratively chosen sequences of thresholds (e.g. TRACE3D, TRT, TITAN).
%Cell identification is generally done by identifying radar regions above a single reflectivity threshold, however some systems use multiple thresholds (e.g. SCIT) or a sequence of thresholds (e.g. TRACE3D, TRT, TITAN). 
The advantage of multiple thresholds is that clusters of storm cells can be identified. The tracking procedure then involves extrapolating the position of cells found in a previous scan (`parent cell') to an estimated new position and checking if there is a match with cells found in the new scan (`children'). The matching is usually done by distance, but the CELLTRACK algorithm also accounts for `shape similarity'.

%How information from multiple previous time steps is incorporated varies between methods. 
Various methods differ in how they incorporate information from multiple previous time steps.
For example, SCIT performs a least-squares fit, but TRACE3D takes a running weighted sum with the oldest information having the least weight. The original TITAN algorithm \citep{DixonM.andWeiner1993TITAN:Methodology} uses an optimisation method for cell tracking. Here all potential tracks are considered and a cost function based upon position difference and volume difference is minimised to select the most likely track. This method is based upon several assumptions, namely that cells move small distances between radar scans and retain their characteristics such as size and shape. Neither of these assumptions are satisfied even approximately and modifications have been made in \cite{Munoz2018EnhancedCells} to incorporate information from an optical flow based field tracker to constrain the solutions from the original method. The CELLTRACK algorithm also uses information from a field-based method (COTREC \citep{Novak2007TheSystem}) to provide a first guess for the new position estimates of cells during tracking.  

Besides allowing for differential motion between nearby cells, cell-based approaches also enable the tracking and prediction of cell characteristics and life cycles. We will return to this idea in section \ref{conv_dev}.

\subsection{Summary}

Lagrangian persistence provides a foundation for most present-day nowcasting systems. Its use involves first estimating a motion field from a sequence of observations, then advecting the most recent observations along this motion field. The motion field is estimated using a variational minimisation or cell-tracking algorithm. 

While Lagrangian persistence provides a useful foundation, the assumption is frequently flouted by real precipitation fields. The following sections will describe extensions to this simple model which loosen the unrealistic assumption of persistence.

\section{Probabilistic and stochastic approaches} \label{prob}

According to \cite{Bowler2006STEPS:NWP}, the most significant sources of error for advection models are:
\begin{enumerate}
\itemsep0em 
  \item errors in estimating the advection field
  \item neglect of temporal evolution of the advection field
  \item neglect of Lagrangian evolution
\end{enumerate}
\noindent with the latter found to be the most significant source of model error.

There are two, non-exclusive, approaches to mitigate these sources of error. The first is to model the uncertainty due to these errors using a probabilistic approach. The second, applicable to the latter two points, is to attempt to improve the model by representing the neglected evolution processes. In this section we discuss probabilistic approaches, with evolution approaches being discussed in section \ref{conv_dev}.

In its most basic form, advective nowcasting is entirely deterministic. However, the importance of accounting for uncertainty of predictions is now widely recognised \citep{Gneiting2008Editorial:Forecasting}, especially in the case of precipitation, where forecast errors can have a significant effect on downstream applications \citep{Vivoni2007ErrorModel}. A number of approaches have therefore been proposed to account for the sources of error listed above, both probabilistic and stochastic.

\subsection{Neighbourhood methods} \label{neighbourhood}
A practical approach to estimating the uncertainty due to errors in the advection field is to consider the 
distribution of precipitation values in the area surrounding the target grid cell. In 
\cite{Andersson1991ARadar}, the neighbourhood considered is a circle with radius proportional to the 
forecast time duration, from which precipitation values are drawn uniformly. 
\cite{Schmid2002Short-termRainfall} assume instead that advection errors follow a two-dimensional 
Gaussian distribution, with lengthscale proportional to the extrapolation time. \cite{Germann2004ScaleForecasts} provide a more detailed analysis of the optimal neighbourhood size as a function of time.
%A more detailed analysis of the optimal neighbourhood size as a function of time is given by \cite{Germann2004ScaleForecasts}.

A related approach is taken by \cite{Fox2005AScheme} and \cite{Xu2005AReflectivities}, which frames the problem as a stochastic linear integro-difference equation (IDE) governed by a redistribution kernel $k$. The redistribution kernel plays a similar role to the neighbourhood, determining the motion and diffusion properties of the field. By allowing $k$ to vary as a function of space, this model can achieve considerable flexibility. Inference, carried out using Markov chain Monte Carlo (MCMC), is correspondingly more intensive.

\subsection{Stochastic methods (Scale decomposition)} \label{decomp}
The persistence time of precipitation features is known to be related to their spatial scale, with smaller 
features having shorter lifetimes \citep{Germann2002Scale-DependenceMethodology}. It is sometimes seen as desirable to filter out these unpredictable scales in order to reduce the root-mean-squared forecast error \citep{Turner2004PredictabilityMAPLE,Seed2003AForecasting}. An alternative to filtering unpredictable lengthscales is to model them stochastically, by injecting spatially-correlated noise. 
%While only the latter approach can be described as probabilistic, the methods are similar enough that we will treat them together.

Both filtering and stochastic simulation rely on an initial scale decomposition step. The simplest approach is to use the Fourier transform to move the field into spectral space, and then apply a Gaussian bandpass filter to divide it into frequency bands \citep{Seed2003AForecasting,Bowler2006STEPS:NWP}. The full field can then be expressed as a sum of these bands:
$$
\Psi_{ij} = \sum_{k=0}^{K-1} \Psi_{kij},
$$
\noindent where $K$ is the number of frequency levels. Since rainfall fields can be highly non-stationary, localised techniques such as the wavelet transform \citep{Turner2004PredictabilityMAPLE} or short-space Fourier transform \citep{Nerini2017ATransform} are sometimes preferred over standard Fourier methods.

When performing the spectral decomposition, it is standard to work with logarithmic radar reflectivity (dBZ) rather than the precipitation rate \citep{Germann2002Scale-DependenceMethodology}, as rain 
rates are believed to be better represented by a multiplicative cascade model than an additive model, in a similar manner to turbulence \citep{Schertzer1987PhysicalProcesses}. The use of a logarithmic variable means that this can be rewritten as an additive process. 

In the case of filtering, high-frequency components are simply truncated as they reach their limits of 
predictability \citep{Seed2003AForecasting}. In stochastic methods, they are instead replaced by 
spatially correlated noise. This is done by convolving a Gaussian white noise field with a filter which captures the correlation properties. In earlier work such as \cite{Bowler2006STEPS:NWP}, the filter is given by a theoretical power-law model; further details are given in \cite{Schertzer1987PhysicalProcesses} and \cite{DApuzzo2008AGenerator}. An alternative, used more recently in \cite{Seed2013FormulationScheme}, is to use the Fourier transform of an observed rainfall field as a filter. Unlike the power-law method this approach is not limited to isotropic fields, and can be used to model noise with a strong directional element. Both approaches are limited to modelling stationary fields, and tend to produce unrealistic levels of noise in areas of high rainfall. However, the use of a logarithmic model does eliminate noise in dry areas.

In the STEPS system, both the precipitation field and the stochastic noise field are evolved 
using a hierarchy of second-order auto-regressive (AR-2) processes \citep{Seed2013FormulationScheme}. The coefficients of these processes determine the rate of Lagrangian evolution for a given spectral band.

\subsection{Summary}

In their most basic form, radar extrapolation approaches to nowcasting assume that the evolution of a 
precipitation field is primarily due to Lagrangian advection. Probabilistic methods relax this 
assumption by accounting for deviations from pure advection, either by interpreting the forecasts as 
neighbourhood rather than point predictions, or by removing the dependence on ``unpredictable'' scales by blurring them or replacing them with random noise. 

The techniques described in this section do allow for some uncertainty in the development of the field due to Lagrangian evolution or variability in the advection field. However, these effects are limited to the 
inherent uncertainty within a given weather regime. They do not account for the more challenging problem of regime change, nor do they account for the possibility of predictable Lagrangian evolution. We will turn to these considerations in the following section.

\section{Nowcasting convective development}
\label{conv_dev}

One problem that has not been addressed by radar extrapolation techniques is that of predicting Lagrangian evolution, the largest remaining source of error in nowcasting. The absence is most notable in the case of convective initiation, development and decay. While these effects can be resolved by high-resolution NWP models, it is not straightforward to combine this predictive ability with the greater accuracy of radar extrapolation methods over short time periods. However, certain indicators have been found to improve nowcasting skill in handling more complex temporal evolution.

Most effort in predicting convection in nowcasting systems has been directed towards deep cumulus convection, because of its association with extreme rainfall. Three main approaches have been used in the literature to predict such events.
\begin{enumerate}
\item Combining advective nowcasting with an NWP model by blending or incorporating specific analyses.
\item An object-oriented approach, where deepening cumulus clouds are detected and classified according to their \textit{development potential}. \citep{Hand1996AnThunderstorms}
\item The detection of boundary layer convergence. \citep{Mueller2003NCARSystem}
\end{enumerate}

A primary finding of the comparison study \cite{Wilson2004SydneyNowcasting} was the significance of the third method. The authors state that \textit{``skill above extrapolation occurs when boundary layer convergence lines can be identified and utilized by a nowcasting system to nowcast storm evolution"}. The Auto-Nowcast system described in \cite{Mueller2003NCARSystem} uses radar to detect and track boundary layer convergence lines. Satellite data has also been used to detect boundary layer convergence, even in the absence of clouds \citep{Roberts2003NowcastingData}, as well as early cumulus growth. However, such signals will not lead to strong convection except in the presence of sufficient instability and moisture. 

In response to this issue, several studies have investigated combining information on potential convection (nascent clouds or convergence) with information on whether the environmental conditions favour convective development. For example, advected satellite imagery can be used to identify potential convective cells, together with relevant NWP analysis fields such as convective available potential energy (CAPE) and convective inhibition (CIN) to predict further development, as in \cite{Steinheimer2007ImprovedFields} and \cite{Mecikalski2015ProbabilisticData}. The general framework of combining advection with diagnoses for initiation allows a great deal of flexibility in the choice of diagnosis, and various data sources and algorithms have been considered. \cite{Ahijevych2016ProbabilisticTechnique} incorporates radar reflectivity as well as satellite data, while \cite{Han2017AData} uses only 3-D radar analysis. Most of these works frame strong convection as a binary decision problem, to which they apply various statistical learning techniques such as logistic regression \citep{Mecikalski2015ProbabilisticData}, random forests \citep{Ahijevych2016ProbabilisticTechnique} and support vector machines (SVMs) \citep{Han2017AData}. There is clearly scope for incorporating ML algorithms of arbitrary complexity within this framework, and we return to this theme in section \ref{sourcespred}.

In cell-based models, various methods can be used to extrapolate from the current characteristics of a cell to its future evolution. A simple example is the TITAN model \citep{DixonM.andWeiner1993TITAN:Methodology}, which assumes that growth and dissipation each follow a linear temporal trend. A more complex development model is given in \cite{Hand1996AnThunderstorms} using a conceptual ``flowchart" model of storm evolution. \cite{Mueller2003NCARSystem} combines a cell-based prediction scheme with additional predictors such as boundary layer convergence lines to more skilfully forecast storm evolution.

The Lagrangian evolution of convective systems is also closely connected to convective initiation. 
When the downdrafts of large convective cells reach the surface, they form an outward-spreading `cold pool' of air. The cold pool can then act to lift the surrounding area, and in favourable conditions this leads to further secondary convection. In the case of multi-cell storms, secondary convection is what enables a storm to survive beyond the first hour or so \citep{Bennett2006AKingdom}. It is also a significant cause of non-advective evolution. To capture these secondary effects, \cite{Hand1996AnThunderstorms} use an object-based system in which mature cumulonimbus clouds initiate daughter cells in their expected downdraft location. This allows the system to predict some nonlinear evolution, although it does not address the initiation of first-generation cells.

\subsection{Summary}

Where it is possible at all, the prediction of Lagrangian evolution depends on incorporating additional diagnostics which give an indication of the potential for convective development. These include: CAPE and CIN; boundary layer convergence; orography; large-scale tropospheric subsidence; moisture; and cold pools due to cell downdrafts. In some cases, it may be possible to infer some of these diagnostics from observed precipitation fields. Others will require extra sources of information such as NWP analyses.

\section{Machine learning approaches}

While methods for radar extrapolation based on optical flow have seen considerable success, they have certain limitations due to assumptions of Lagrangian persistence and smooth motion fields. There have been efforts to relax these assumptions by incorporating specific mechanisms (such as convergence lines or cell life-cycles), or via probabilistic approaches; but as yet there has been no fully general solution short of a costly data assimilation cycle. Researchers have begun to explore machine learning techniques as a way to fill this gap. Machine learning provides an opportunity to capture complex non-linear spatio-temporal patterns and to combine heterogeneous data sources for use in prediction. In principle, it can enable us to weaken the assumption of Lagrangian persistence and produce more flexible models which take advantage of more varied sources of predictability. 

Nonetheless, several challenges arise in adapting well-developed tools from deep learning to nowcasting. Having originally been developed to address quite different problems, even reproducing flow-based methods in a deep learning context can be challenging. 
Furthermore, the expressivity of deep learning models can produce results with undesirable properties such as blurred fields and missed extreme events if optimising with standard metrics such as mean-squared errors (MSE).
%What's more, the flexibility of deep learning means that optimising against standard metrics such as mean-squared errors (MSE) can produce results with undesirable properties such as blurred fields and missed extreme events. 
Finally, interpretability remains a challenge. The main benefit of deep learning lies in its flexibility to learn arbitrary functions, but this very flexibility makes it more difficult to properly analyse the contribution of a model, and to understand which features are essential to its performance and which are incidental. 

This section describes machine learning approaches that have already been applied to nowcasting, and also describes some techniques for related problems we believe could be fruitfully applied in this field.

\subsection{Dense spatiotemporal prediction}
\label{dense}

In NWP there exists a well-developed paradigm for dense spatiotemporal prediction. In ML, the situation is still developing: there are numerous possible approaches, and not necessarily clear criteria for which approach is more appropriate for a particular task. This section describes three different options, based on recurrent convolution, flows, and direct prediction.

\subsubsection{Spatiotemporal convolution}
\label{conv}

A common approach to temporal modelling is to use recurrent neural networks (RNNs) as they are designed to model an evolving state over time. This works by using the current network state as input into the next timestep. More recent variants such as long-short term memory (LSTM) \citep{Hochreiter1997LONGMEMORY} and gated recurrent unit (GRU) \citep{Cho2014OnApproaches} have incorporated gating structures to protect information over time, making it possible to learn long-range temporal dependencies. They have been applied to video prediction in \cite{Oh2015Action-ConditionalGames}, which combines a convolutional encoder-decoder architecture with a fully-connected LSTM operating on the latent feature vector. This method was adapted by \cite{Shi2015ConvolutionalNowcasting} to retain spatial structure in the latent representations, enabling them to use convolution operations for LSTM state-state transitions instead of a fully connected layer.

Besides incorporating recurrence, the time dimension can also be handled as part of a convolutional architecture. One way to achieve this is by encoding the sequence of frames along the channel dimension and performing 2D-convolution \citep{Mathieu2015DeepError}. Another approach is 3D-convolutions, which are used by \cite{Vondrick2016GeneratingDynamics} in the context of video generation. For prediction tasks, as future frames can depend only on past frames, a more suitable variant is causal convolutions \citep{Oord2016WaveNet:Audio}, which have been used for video prediction as part of a fully convolutional model in \cite{Xu2018PredCNN:Convolutions}.

At present, it is not clear whether combined convolutional-recurrent architectures or pure convolutional 
architectures are more suitable for spatiotemporal prediction. While there is evidence that well-designed 
convolutional architectures can outperform recurrent architectures for pure sequence forecasting 
\citep{Bai2018AnModeling}, the comparison in \cite{Shi2017DeepModel} suggests that this may not hold true in the spatiotemporal case. A detailed comparison of fully-convolutional and mixed recurrent-convolutional architectures for spatiotemporal prediction has yet to be carried out, and may well prove valuable.

\subsubsection{Flows and deformations}
\label{flow}

Standard convolutional architectures learn a single set of filters that is then applied to every input. This is suboptimal for tasks such as nowcasting, because the appropriate transformation of the input is highly dependent on the input itself, and is also location-varying. Several works address this by using a two-part architecture, with one part predicting an appropriate set of filters and the other applying these filters to the input. 

In \cite{Klein2015APrediction}, this kind of two-part architecture was applied to a radar extrapolation task. Similar to \cite{Austin1978TheProject}, this model learns one or more filters that are applied uniformly to the input; that is, the deformation varies by input, but not by location. Later studies extended this architecture to non-stationary deformations, either by learning a spatially varying filter as in \cite{DeBrabandere2016DynamicNetworks}, or by learning several filters together with masks defining their domain of application, as in \cite{Finn2016UnsupervisedPrediction}. A similar non-stationary structure is used in \cite{Shi2017DeepModel}. However, in this case the model does not predict a convolutional filter, but instead predicts the recurrent connection structure, meaning it is less constrained by the convolution size. 

As pointed out in \cite{Shi2018MachineSurvey}, these approaches are similar in spirit to predicting a spatial deformation field, and therefore comparable to the methods outlined in section \ref{advection}. The work \cite{de_B_zenac_2019} on sea surface temperature prediction makes the use of flow fields more explicit. They use a network conditioned on previous time-steps to predict a motion field, which is then used as input to a warping scheme to transform the input field. The warping scheme computes a weighted average over a Gaussian centred at the previous location of each target cell; essentially a backwards semi-Lagrangian scheme described in section \ref{num_advect}. This model is therefore closely connected to those discussed in section \ref{est_advect} that aimed to estimate a motion field by minimising Lagrangian evolution in the recent past. However, unlike those models there is no explicit minimisation of Lagrangian evolution, since the model is trained end-to-end based on the prediction accuracy. In principle, this gives the model greater flexibility in situations where Lagrangian evolution is not negligible.

More generally, there have been efforts towards training neural networks within a differential equation structure to model physical systems, without explicitly incorporating spatial flow. The authors of \cite{ayed2019learning} demonstrate how this approach can be used to predict the future state of the shallow water model. Unlike the \cite{de_B_zenac_2019} model, this model has freedom to learn an appropriate warping scheme rather than just the motion field.

\subsubsection{Direct prediction and training strategy}

Besides recurrent and flow-based models, it is also possible to treat nowcasting as a direct image-to-image translation problem. This is the approach taken by \cite{agrawal2019machine}, who train a U-Net directly on one-hour prediction of radar fields. 

In general, direct prediction may have an advantage over iterative prediction by allowing the model to tailor its predictions to the target forecast period. This is discussed in \cite{Shi2018MachineSurvey}, where it is argued that the choice of a training strategy is important for sequence forecasting as well as the choice of architecture. It concludes that direct prediction can lead to greater accuracy over short time periods, while iterative prediction may suffer from accumulated errors due to the discrepancy between its training and test objectives. On the other hand, iterative models are easier to train and can be used to generate predictions at any distance into the future, while direct prediction is restricted to the time range it has been trained against. A potential synthesis can be found in the boosting strategy of \cite{Taieb2014BoostingForecasts}, which uses a core iteratively trained model with directly trained adjustments. The recent work of \cite{snderby2020metnet} also combines the two strategies by conditioning an iterative model on the target lead time.

\subsection{Sharp predictions and uncertainty}
\label{uncertain}
In machine learning, spatiotemporal sequence forecasting has most often been treated as a deterministic problem. As observed by \cite{Mathieu2015DeepError}, this approach can have unwanted consequences if the underlying system is not deterministic. If the distribution is multi-modal, attempting to minimise the mean squared error loss will tend to average over the different modes, leading to blurry predictions. 

The alternative is to use a generative approach, sampling from the distribution of possible futures instead of returning a single ``best" prediction. In the spatiotemporal case, this distribution will generally be complex and high-dimensional, making it difficult to model. Even so, a number of methods have been proposed for generative modelling of spatiotemporal sequences. These methods belong to three main categories: adversarial training, sequential conditional pixel generation, and latent random variable models.

%It should be noted that deterministic models will generally outperform generative models on pixel-wise metrics such as MSE. To properly verify these models, it is necessary to include spatial and probabilistic verification metrics in the analysis. While a detailed discussion of verification is outside the scope of this review, reviews such as \cite{Gilleland2009IntercomparisonMethods} and \cite{Dorninger2018TheProject} can provide an entry point.

\subsubsection{Adversarial training}
As proposed in \cite{Goodfellow2014GenerativeNets}, generative adversarial networks (GANs) consist of a pair of networks which are trained in an alternating fashion. One of these networks (the generator) attempts to generate samples from the training distribution, and the other (the discriminator) tries to distinguish between the generated samples and training data. Adversarial training is equivalent to minimising the divergence between the generated and target distributions, and can be viewed as constructing a learned loss function which targets the task at hand, embodied in the discriminator. For prediction tasks, it is essential to be able to condition the generated output on previously observed frames; this can be achieved by providing the conditioning information as input to both the generator and the discriminator \citep{Goodfellow2014GenerativeNets}.

Several studies have considered adversarial training for video generation 
\citep{Vondrick2016GeneratingDynamics,Denton2015DeepNetworks} and video prediction 
\citep{Mathieu2015DeepError,Jang2018VideoConditions}. In particular, \cite{Mathieu2015DeepError} found that an adversarial training objective led to consistently sharper predictions than either the $\ell_2$ or $\ell_1$ norm. 

Beyond producing sharp predictions, we may hope that the distribution of samples produced by a GAN is close to the true distribution. This is known to be the case in the idealised scenario of unlimited network capacity and training data, provided the model converges \citep{Goodfellow2014GenerativeNets}. However, in more realistic scenarios convergence does not imply that the true distribution has been well approximated \citep{Arora2017GeneralizationGANs}, and in practice GANs often suffer from mode collapse \citep{Santurkar2017ADistributions}. Evaluating high dimensional generative models is a significant open problem \citep{Gulrajani2019TowardsGeneralization}. These issues are especially acute for GANs because 
they do not explicitly model a probability distribution.

\subsubsection{Sequential conditional pixel generation}
Besides adversarial methods, other techniques aim to specify the full conditional distribution over 
predictions. One approach, described in \cite{Kalchbrenner2016VideoNetworks}, is to factorise the 
likelihood as the product of conditional probabilities over pixels, so that each pixel is conditioned on 
previous timesteps and on previously generated pixels in the current timestep (in their implementation, the pixels above and to the left). This conditioning ensures that the sampling of all pixels is consistent, in that they are all drawn from the same mode of the full distribution. The only requirement is that the model for the conditional distributions should be flexible enough to capture the dependencies (for example, it should not be assumed to be unimodal) while still being simple enough to provide tractable inference. In \cite{Kalchbrenner2016VideoNetworks}, the authors choose a discrete distribution over 256 potential values, essentially modelling the distribution by a histogram.

It is worth noting that \cite{Kalchbrenner2016VideoNetworks} also model conditional dependencies across colour channels by imposing an arbitrary order. Although radar-based nowcasting is usually limited to a single channel, the same technique could be useful for modelling dependencies between meteorological variables (e.g. temperature and rainfall), or between radar and satellite observations.

In principle, the value of a pixel can be conditioned on all previous timesteps and pixels using a 
gated RNN, as described in \cite{Oord2016PixelNetworks}. An alternative option is to condition on only a bounded region of pixels, which can be done using a CNN. This approach makes training significantly 
faster, since the convolutions can be run in parallel; the downside is that it introduces artificial conditional independence assumptions \citep{Oord2016ConditionalDecoders}. The video prediction model in \cite{Kalchbrenner2016VideoNetworks} uses a hybrid approach, namely convolution in the spatial dimensions and a convolutional LSTM over time. 

In either case, whether using the RNN or CNN version, using the model to generate samples requires 
generating every pixel sequentially. As a result, this class of models is time consuming for spatiotemporal data, especially at high resolutions.

\subsubsection{Latent random variable models}

Most deep learning research on latent random variable models focuses on variational autoencoders (VAEs). VAE is an encoder-decoder architecture which attempts to learn an embedding of the observed fields into a latent space. The model is trained by approximate inference, as detailed in 
\cite{Kingma2014Auto-EncodingBayes}. The training objective is essentially to minimise the reconstructive error, but with an extra regularisation term which seeks to minimise the divergence between the encoded latent distribution and some fixed prior distribution. This prior is often taken to be a Gaussian with a diagonal covariance matrix, but this is not essential. What matters is that the prior distribution is known and can be sampled; by regularising the encoded random variables so that they are close to the prior, it is then possible to use samples from the prior to generate samples from the target distribution by passing them through the decoder.

Several works have incorporated a VAE component in their architecture to create a probabilistic model, 
such as for image prediction \citep{Xue2016VisualNetworks} and segmentation \citep{Kohl2018AImages}. In the context of video prediction, both \cite{Babaeizadeh2017StochasticPrediction} and \cite{Denton2018StochasticPrior} have proposed combining a VAE with a spatiotemporal LSTM predictor. However, the models differ in their treatment of the latent random variables: while \cite{Babaeizadeh2017StochasticPrediction} uses samples from a constant distribution to capture the system stochasticity, the model of \cite{Denton2018StochasticPrior} learns to predict a distinct prior at every timestep. The time-varying prior is similar to that used by \cite{Chung2015AData}, and was found to significantly improve the sharpness of predictions. It is also intuitively a better model of nonstationary environments, since the form of predictive uncertainty could change entirely from one frame to another.

A related group of generative models, designed to perform approximate inference, is flow-based models 
(NICE \citep{Dinh2014NICE:Estimation}; RealNVP \citep{Dinh2016DensityNVP}). Instead of using a 
variational autoencoder to train the parameters of the latent distribution, these models use invertible neural networks, which enable direct optimisation of the log likelihood. The idea is to approximate complex, highly non-Gaussian posteriors using a series of invertible transformations; this technique is described in \cite{Rezende2015VariationalFlows} as variational inference with normalising flows. Flow-based models have been applied to conditional video prediction by \cite{Kumar2019VideoFlow:Video}. In this work, invertible networks are used to connect the observed and latent variables at each timestep and level of detail, while non-invertible networks are used to model the conditional dependencies between timesteps and levels.

\subsection{Additional sources of predictability}
\label{sourcespred}

Due to their flexibility as universal function approximators, machine learning methods could be expected to tease out additional sources of predictability implicit in the nowcasting problem, beyond that exploited by extrapolation. One example is that of accounting for uncertainty with blurred predictions. Others include: 

\begin{enumerate}
\item influence of orography
\item life-cycle of convective cells
\item predictable motion field evolution
\item convergence lines
\item stable and unstable regimes.
\end{enumerate}

\noindent Each is a potential source of predictability which could lead to improved skill compared to pure optical flow. In order to realise this additional skill, an ML model needs to have access to the right information. This could be provided to the model in the form of additional predictors, or in some cases extracted by the model itself from information already implicit in the radar data. For example, orographic fields could be used as input to an ML model, or the model could infer this information from raw location data. Likewise, convergence lines could be detected by adding assimilated wind fields as an input, but could also potentially be inferred from spatial patterns in the radar data using a convolutional neural network (CNN).

The example that has so far received most attention is incorporating orographic data. \cite{Foresti_2019} found that a neural network trained on spatial location and flow direction had significantly better skill at predicting growth and decay of precipitation than a simple Eulerian persistence model, with inclusion of the previous rainfall value improving performance further. The effect was particularly strong over mountainous areas. Another example is found in \cite{Franch_2020}, where orographic data was used as input to a post-processing model stacked on top of a deep learning ensemble to improve predictions of extreme rainfall.

Motion field evolution is a promising target for ML because, unlike standard optical flow, most of the models discussed in sections \ref{conv} and \ref{flow} are able to represent time-varying motion fields. While not itself a machine learning model, \cite{RyuImprovedEquation} gives an encouraging signal that the evolution of motion fields may have some predictability without incorporating a full NWP model; they found that evolving the motion field based on Burgers' equation improved CSI compared to assuming a constant motion field.

As outlined in section \ref{conv_dev}, machine learning techniques have also been used to predict strong convective initiation by incorporating various analyses. Another attempt in this direction is \cite{HanSVM}, in which a support vector machine (SVM) is trained to predict whether a box will contain high rainfall 30 minutes into the future based on vertical velocity and buoyancy values in nine neighbouring boxes. The input data is based on high-resolution VDRAS analysis \citep{SunVDRAS}. It is argued that the SVM can predict storm growth and convective initiation, although this may be an artefact due to the over-forecasting bias shown by the model.

So far, most research into sources of predictability has either treated the problem in isolation (as in the case of \cite{Foresti_2019} or used the predictions to augment simple radar extrapolation. There has not yet been a study of how these sources of predictability are utilised by more complex deep learning architectures such as those described in sections \ref{dense} and \ref{flow}. In part, this is no doubt due to interpretable machine learning \citep{McGovernInterpretability} being a challenging area in its own right. Perhaps more significantly, such process-oriented analysis requires an understanding of the mechanisms at play in meteorology and atmospheric science, as well as expertise in ML: further motivation for bridging the gap between the two fields.

\subsection{Summary} \label{mlsummary}

There has been considerable progress in adapting deep learning methodology to dense spatiotemporal prediction tasks. Nowcasting itself has been addressed by direct prediction, as well as composite architectures combining recurrent and convolutional connections, and others using temporal convolutions. There have also been efforts to account for differing background flows by making the connection structure itself a predictand. 

In other contexts, deep learning models have been used to predict flow fields that are used to transform the input, with gradients back-propagated through the flow. Although these methods have not been specifically applied to nowcasting there is a clear connection with the advection models discussed in section \ref{advection}, and comparing these approaches could be an interesting direction for future research.

Another area with considerable scope for development is handling the multimodal system dynamics which occur in nowcasting. This will require using a generative approach, such as adversarial training, sequential pixel prediction, or variational auto-encoders. Each of these methods has been applied with considerable success in video prediction and generation; in particular, using a variational auto-encoder with a distinct prior predicted at each time-step seems a promising approach for atmospheric science \citep{Chung2015AData}. 

A key advantage of machine learning models is their flexibility in combining data sources and utilising more varied sources of predictability. Despite this potential, so far little attention has been paid to combining multiple data sources, with the exception of orographic data which has been shown to hold predictive skill \citep{Foresti_2019}. It also remains to be shown whether deep learning architectures such as CNNs are able to extract additional predictability from the spatial structure of input fields.

\section{Discussion and Conclusion }

The nowcasting problem poses several distinct research challenges: 
the need to make dense
predictions; the need to handle underlying structure at multiple spatial and
temporal scales; the need to handle extremes and out-of-sample events; an underlying
need for flow prediction and handling of regime changes; and the need to make and verify
 probabilistic predictions. This review identifies some methods that may help solve those challenges.
 
%Individually, these are important research questions in their own right, but collectively, and for the success of nowcasting, require new approaches.  This review identifies some methods that may help solve those challenges.
%More importantly, however, the combination of atmospheric science and machine learning, which have been looked at individually in this paper, will make possible the development of such new solutions.

Making the best possible use of other observational data to aid short-term
rainfall prediction remains an active and challenging research area. There is
opportunity to make progress in this area by fusing machine learning and
meteorology; machine learning for its ability to extract meaning from
high-dimensional data, and meteorology for its elucidation of the underlying
physics driving evolution.

%A clear priority for future research is predicting convective initiation andevolution. This is an area where extrapolation approaches have limited ability,although they have seen benefits from including additional information on areasof convective potential. Similarly, incorporating additional fields is apossibility for improving machine learning models. Determining which fieldsprovide the most value will require further investigation, but CAPE and boundarylayer convergence seem to be promising candidates. 

Incorporating physical mechanisms into ML models is a promising direction for further research, that may help to improve predictions and interpretability. This is especially true at longer time scales, where understanding the physics of the full atmosphere becomes important for understanding the evolution of the rainfall field. Physics-based advection models would be one example of this type of hybrid approach, as would modelling the effects of orography. Predicting convective initiation and evolution is a clear priority for future research in this area.

An advantage of deep learning, specifically CNNs, is their ability to extract latent information from spatial structure. This may enable them to make more efficient use of the data provided to them. For example, it may be possible to infer areas of developing boundary layer convergence from radar data within the model, rather than including it as an additional input. Such models may provide an alternative route to the incorporation of physical mechanisms discussed above; but there is still work to be done investigating the flow of information through these complex models.

One important development in the literature on extrapolation-based nowcasting
has been the move away from optimal mean prediction and towards stochastic and
probabilistic approaches, in order to account for the chaotic nature of
convective evolution. This viewpoint also seems likely to be useful for
approaches based on machine learning. The extensive literature on deep
generative models gives hope that they will help to shed light on the
challenging problem of representing uncertainty of convective-scale evolution,
which is frequently multi-modal and nonlinear.

Finally, any new research will need to consider carefully the question of
verification. A huge number of possible metrics are available, each of which
embodies a particular view of what it means to be a ``good'' forecast. There is
no general consensus on a single best verification metric which should be used,
and such a consensus probably is not possible. The most reasonable approach
seems to be to use a range, including at least one point-based metric, one spatial metric, and one for assessing ensemble calibration where applicable.

The weather touches every part of human life and with the need for greater
resilience to changing environmental conditions the need for more accurate
nowcasting is of increasing importance. There is much opportunity for new
research and we believe that the fusion of the environmental sciences and
machine learning holds much promise.

\bibliography{nowcasting}
\clearpage

\end{document}